# Complementary Goodness of Fit Procedure for Crash Frequency Models

**Mohammadreza Hashemi** [a], **Adrian Ricardo Archilla** [a*]

[a] Dept. of Civil and Environmental Engineering, Univ. of Hawaii at Manoa, Honolulu, HI 96822, United States

## ABSTRACT

This paper presents a new procedure for evaluating the goodness of fit of Generalized Linear Models (GLM) estimated with Roadway Departure (RwD) crash frequency data for the State of Hawaii on two-lane two-way (TLTW) state roads. The procedure is analyzed using ten years of RwD crash data (including all severity levels) and roadway characteristics (e.g., traffic, geometry, and inventory databases) that can be aggregated at the section level.

The three estimation methods evaluated using the proposed procedure include: Negative Binomial (NB), Zero-Inflated Negative Binomial (ZINB), and Generalized Linear Mixed Model-Negative Binomial (GLMM-NB). The procedure shows that the three methodologies can provide very good fits in terms of the distributions of crashes within narrow ranges of the predicted mean frequency of crashes and in terms of observed vs. predicted average crash frequencies for those data segments.

The proposed procedure complements other statistics such as Akaike Information Criterion, Bayesian Information Criterion, and Log-likelihood used for model selection. It is consistent with those statistics for models without random effects, but it diverges for GLMM-NB models. The procedure can aid model selection by providing a clear visualization of the fit of crash frequency models and allowing the computation of a pseudo $R^2$ similar the one used in linear regression.

It is recommended to evaluate its use for evaluating the trade-off between the number of random effects in GLMM-NB models and their goodness of fit using more appropriate datasets that do not lead to convergence problems.

## INTRODUCTION

Highway fatalities and severe injuries in the United States continue to be a major national concern. Although there has been a steady downward trend in both the fatal crash rates and the total number of traffic fatalities, in 2017 there were still 34,247 annual fatalities which indicates that much work is still required (*1, 2*).

There has been steady progress in developing mathematical approaches to capture the complexity of crashes (*3*). Two types of models of interest are crash frequency models and crash severity models. This paper is concerned with the goodness of fit evaluation of the former.

Different methodologies have been applied in crash frequency analyses attempting to capture the data generation process. Such methodologies include generalized linear models (GLM) using different distributions such as Poisson (*4*), negative binomial (*5*), zero-inflated negative binomial (*6*) and other distributions (*7, 8, 9*). Furthermore, various statistical approaches such as random parameter negative binomial (*10*) and mixed-effects negative binomial models (*11*) have been introduced to consider the unobserved heterogeneity arising from the heterogeneous effects of roadway characteristics across the observations due to the unobserved time-varying environmental conditions as well as the heterogeneous reaction of drivers to the roadway features (*9, 12*).

---
[*] Corresponding author
E-mail addresses: hashemi@hawaii.edu (M. Hashemi), archilla@hawaii.edu (A.R. Archilla)

While different statistics are used to evaluate the quality of their estimation, there is no equivalent to an observed vs. predicted graph that allows for an easy understanding of the goodness of fit of the model. Furthermore, the usefulness of a plot of residuals or deviances in GLMs is often limited.

This paper proposes a simple procedure to compare the distributions of predicted and observed crash rates for GLMs that accounts for the probabilistic nature of the modeling and allows the creation of a graph of observed vs. predicted average crash frequencies and the estimation of a pseudo $R^2$. It is not intended to substitute other widely used statistics such as the Akaike Information Criterion (AIC), the Bayesian Information Criterion (BIC), or likelihood ratio tests used to select competing models but instead, to complement them by providing a more intuitive way to visualize the models' goodness fit.

The procedure is presented with models of roadway departure (RwD) crash frequency on sections of two-lane two-way (TLTW) state roads developed with RwD crash data from the State of Hawaii. The models are based mostly on roadway characteristics and on RwD crashes without regard for the severity since there are very few fatalities and severe injury crashes per year in Hawaii.

The focus of this paper is not in the model estimation but on its goodness of fit, consequently, the data and model features are introduced only briefly for completeness. Further model details and insights toward understanding the contributing factors of RwD crashes on TLTW state roads in Hawaii, the potential use of the models for identification of segments with a higher risk of RwD crashes and selection of safety countermeasures to alleviate the RwD crashes are left for discussion in separate papers.

**OBJECTIVE**
The objective of this paper is to present the rationale for the development of graphs and associated goodness of fit procedure applicable to crash frequency models that complements commonly used statistics for selecting and evaluating generalized linear models.

## BACKGROUND
The total number of crashes observed on roadway sections is an example of count data, with only non-negative integer outcomes. Count data is properly modeled by GLMs such as Poisson and negative binomial (NB) regression models since they are suitable for predicting the probability of observing rare events modeled by non-negative integers (*13*).

Lord and Mannering reviewed different methodological alternatives for modeling crash frequencies (*14*), including Poisson, Negative Binomial (NB), and Zero-Inflated Negative Binomial (ZINB) models (among others). A significant limitation of Poisson regression is its requirement that the mean of the count data be equal to its variance. When the variance is larger than the mean, the data are said to be over-dispersed. For such data, a NB model is more appropriate (*15*). Consequently, only the NB model together with its ZINB and Generalized Linear Mixed Model – Negative Binomial (GLMM-NB) variations are briefly summarized below.

### Negative Binomial (NB) Regression
As in Poisson regression, in NB regression, the expected crash frequency for segment $i$ is modeled as a function of explanatory variables collected in a vector $\boldsymbol{x}_i$ such that:

$$\lambda_i = ex\,p(\mu_i) = exp\,(\boldsymbol{x}_i\boldsymbol{\beta} + \varepsilon_i) \tag{1}$$

where
$\lambda_i =$ expected crash frequency for section $i$,
$\boldsymbol{x}_i =$ row vector of explanatory variables for section $i$,
$\boldsymbol{\beta} =$ column vector of estimable parameters, and
$\varepsilon_i =$ an error term for section $i$.

One of the parameterizations of the negative binomial distribution for predicting the probability of observing $n_i$ crashes on section $i$ is:

$$P(n_i) = \left(\frac{\Gamma\left(n_i+\frac{1}{\alpha}\right)}{n_i!}\right) \times \left(\frac{\frac{1}{\alpha}}{\frac{1}{\alpha}+\lambda_i}\right)^{\frac{1}{\alpha}} \times \left(\frac{\lambda_i}{\frac{1}{\alpha}+\lambda_i}\right)^{n_i} \tag{2}$$

where $\Gamma(\cdot)$ is the Gamma function and $\alpha$ is the overdispersion parameter. The vector of parameters for this model ($\boldsymbol{\beta}$) can be estimated by maximum likelihood. Winkelmann explains that $\varepsilon_i$ can be considered as a random-effect term to model the individual heterogeneity (*16*). It is commonly assumed that exp ($\varepsilon_i$) is Gamma distributed with a mean of one and a variance of $\alpha$. The addition of the error term $\varepsilon_i$ allows the variance of $n_i$ to differ from the mean:

$$VAR[n_i] = E[n_i]\big[1 + \alpha E[n_i]\big] \tag{3}$$

### Zero-Inflated Negative Binomial Regression (ZINB)
The ZINB model addresses the problem of having a high fraction of zeros in the response variable. Zero events during the observation period can result from two separate representations: one set of observations that are necessarily zeroes and another set generated from a NB process with some additional probability of observing zeros. In terms of crash frequency data analysis (*15*), these representations imply that either the segment is absolutely safe and the probability of crash occurrence is zero, or that the failure to observe any crashes in the period of study comes from the representation with a low probability of observing a zero crash. Following the notation in Winkelman ($16$), a ZINB model combines a binary variable $C_i$ with a standard count variable $n_i^*$ that is given by:

$$n_i = \begin{cases} 0 & if\ C_i = 1 \\ n_i^* & if\ C_i = 0 \end{cases} \tag{4}$$

If the probability that $C_i = 1$ (stating the section is absolutely safe) is denoted by $w_i$, the probability function of $n_i$ is:

$$P(n_i) = w_i d_i + (1 - w_i)g(n_i) \tag{5}$$

where $d_i = 1 - \min\{n_i, 1\}$ and $g(n_i)$ is a NB distribution ($16$). Usually, a binary logit model is used to model the binary status of the two parts of a zero-inflated model. Lord et al. provide further notes on the detailed application of this model for safety data analysis ($17, 18$).

**Generalized Linear Mixed Effects Model-Negative Binomial (GLMM-NB)**
The GLMM-NB formulation considers a fixed parameter NB regression for each explanatory variable across all the observations. However, it is widely accepted that there are many threads of unobserved heterogeneity in crash data analysis that may change the effect of explanatory variables for different observations ($12$). Therefore, to account for the unobserved heterogeneity, it is common to add a random term (i.e., a random effect) to the parameter of each explanatory variable $l$ as follows:

$$\beta_{il} = \beta_l + \varphi_{il} \tag{6}$$

where $\beta_{il}$ is the parameter on the $l^{\text{th}}$ explanatory variable for observation $i$, $\beta_l$ is the mean parameter on the $l^{\text{th}}$ explanatory variable across all the observations (i.e., a fixed effect), and $\varphi_{il}$ is a randomly distributed scalar term that captures unobserved heterogeneity across the observations (i.e., a random effect) with a distribution assumed by the analyst (e.g., normal). Basically, if the variance of the chosen distribution is not significantly different from zero, only the fixed-effect parameter would remain in the model. The likelihood function for this model can be written as:

$$L = \prod_{\forall i} \int_{\varphi_i} P(n_i | \varphi_i)\, g(\varphi_i)d\varphi_i \tag{7}$$

where the $g(\varphi_i)$ is the probability density function of $\varphi_i$.

The approach is known as random parameter NB regression model in the crash frequency modeling literature ($10, 19$), which is equivalent to what is known as Generalized Linear Mixed Effects Model-Negative Binomial (GLMM-NB) in the terminology commonly used in statistics ($11, 20$). Generally, a mixed-model contains both fixed effects and random effects. Fixed effects are parameters which can be associated to the entire population explaining the behavior of the population means whereas random effects are associated with individual sections which are drawn at random from an entire population ($21$). In modeling crash frequencies with a GLMM-NB model, fix-effects describing the average variation of the mean number of crashes on section $i$ are included with a set of covariates gathered in a design matrix $\boldsymbol{X}$, whereas random-effects related to explanatory variables for each section are collected on a matrix $\boldsymbol{Z}$ used to consider the unobserved heterogeneity effects across sections. For identifiability, the mixed-effects model typically requires the availability of either repeated measurement data (such as observing number of crashes on consecutive years and treating these as separate observations) or grouped data to model the response variable as a function of explanatory variables while considering the correlation between observations in each group and the variation between groups that might affect the response variables (*21*). This separation of the two sources of variation results in consistent and efficient variance standard errors and in turn, in more reliable statistical results for the parameter estimates.

**Rate Models**
The total number of observed RwD crashes on a roadway section $i$ depends on the length of the section $L_i$, its Annual Average Daily Traffic ($AADT_i$), and the number of years over which the number of crashes was observed $N_i$. These variables determine the number of opportunities for the event to occur (*11*). This can be incorporated using the following approach. Consider Eq. 1, but written in terms of the individual independent variables:

$$\lambda_i = e^{\beta_0 + \beta_1 x_{i1} + \beta_2 x_{i2} + \ldots + \beta_n x_{in} + \varepsilon_i} \tag{8}$$

where $x_{ij}$ is the $j^{\text{th}}$ variable for section $i$. Letting $\beta_1 = \beta_2 = \beta_3 = 1$, $x_{i1} = \log(AADT_i)$, $x_{i2} = \log(N_i)$, and $x_{i3} = \log(L_i)$ yields

$$\lambda_i = e^{\beta_0 + \log(AADT_i) + \log(N_i) + \log(L_i) + \beta_4 x_{i4} + \ldots + \beta_n x_{in} + \varepsilon_i} \tag{9}$$

Taking the logarithm of both sides results in

$$\log(\lambda_i) = \beta_0 + \log(AADT_i) + \log(N_i) + \log(L_i) + \beta_4 x_{i4} + \cdots + \beta_n x_{in} + \varepsilon_i \tag{10}$$

which can be further rearranged as:

$$\log\left(\frac{\lambda_i}{AADT_i \times N_i \times L_i}\right) = \boldsymbol{x}_i^r \, \boldsymbol{\beta}^r + \varepsilon_i \tag{11}$$

with the reduced vectors $\boldsymbol{\beta}^r = \{\beta_0, \beta_4, \ldots, \beta_n\}'$ and $\boldsymbol{x}_i^r = \{1, x_{i4}, \ldots, x_{in}\}$. Thus, it is seen that estimating a model with the parameters of the logarithm of variables determining the number of opportunities fixed at one as in Eq. (9) is equivalent to estimating a model for the log of the rate ($\lambda_i / (AADT_i \times N_i \times L_i)$) in Eq. (11).

In R (*22*), the software used for model estimation in this study, rate models are easily implementable with the help of the offset command. This command forces the model to assume a fixed coefficient equal to one for each variable specified with an offset command. Modeling the frequency of RwD crashes as a rate model enables the estimation with sections of different lengths or a comparison of estimated parameters of separately developed models with different section lengths. In this study, all the sections had the same length. It also facilitates the comparison of the effect of explanatory variables on the safety of the roads with different traffic levels. Furthermore, it permits the estimation of models with a different number of years of data available for different sections.

When using offset commands, the parameters are estimated using the total number of RwD crashes, but they still correspond to the rate model. It should also be noted that for a model that considers the rate to be dependent on log(AADT), the estimated parameter for log(AADT) ($\beta_j$) (say log(AADT) is the $j$th variable) for a model estimated without an offset for log(AADT) is related to the parameter of log(AADT) ($\beta_j^r$) of a model estimated with an offset for log(AADT) by $\beta_j^r = \beta_j - 1$. In other words, the use of an offset for log(AADT) is simply a convenient way to account for exposure, but this could also be inferred from a model without offsets as long as log(AADT) is one of the independent variables.

## DATA DESCRIPTION AND SYNTHESIS

### Data Sources
The State of Hawaii motor vehicle accident reports for ten consecutive years (2005-2014) were obtained from the Hawaii Department of Transportation (HDOT). These were used to extract the RwD crashes on TLTW state roadways in the islands of Hawaii, Kauai, Maui, and Oahu. Roadway characteristics were also obtained from HDOT and synthesized for each roadway section. The portion of the state highway network selected for the study covers 600 centerline miles (1200 lane-miles) of TLTW roads, which is slightly less than half of the state highway network.

### RwD Crashes
RwD crashes were extracted using the FHWA definition: a non-intersection crash in which a vehicle crosses an edge line, a centerline, or otherwise leaves the traveled way *(1)*. Table 1 presents a list of first harmful events that are considered as resulting in RwD crashes. It also shows the first harmful events differentiating whether a crash is a RwD to the right side or to the left side of the road. To maintain the compatibility with the FHWA RwD crash definition, the locations of all crashes were filtered to the non-intersection crashes.

### Data Synthesis
The models presented in this paper are based on a 0.2-mile fixed section length. This section length was selected as a compromise between the desire to make the sections as short as possible (hence, more homogeneous) but avoiding an excessive number of sections with zero crashes after a detailed study using different fixed-lengths (0.1-, 0.2-, 0.3-, 0.5-, 1.0-, and 2.0-mile section lengths). Further details of this analysis will be presented elsewhere.

A total of 4,604 RwD crashes were observed on TLTW state roads during the study period (i.e., 2005-2014). Figure 1 presents the distribution of RwD crashes based on their severity according to Hawaii's police crash reports. The figure shows that the total number of RwD crashes with severe injuries (fatal and incapacitating injuries) are very limited. Therefore, all the RwD crashes without regards for the severity are used to develop the crash frequency models. Furthermore, for most sections, the annual number of crashes are either zero or one. Hence, the models were developed with the number of crashes in the 10 years as a single observation in the section, instead of considering the data for each year as a separate observation.

**TABLE 1  List of first harmful events resulting in RwD crashes**

| Type of Collision | RwD to the right-side | RwD to the left-side |
|---|---|---|
| **Non-Collision** | 02 Overturn/Rollover Off Roadway<br>03 Submersion<br>06 Ran Off Roadway | 11 Cross Median/Centerline |
| **Collision with Object/Animal** | 21 Guardrail Face, 22 Guardrail End, 23 Culvert, 24 Ditch, 26 Bridge Pier or Support, 27 Bridge Rail, 28 Building, 29 Tunnel, 30 Curb, 31 Embankment /Retaining Wall, 32 Fence, 33 Utility Pole/Light Support, 34 Traffic Signal/Sign Post, 35 Other Post/Pole/Support, 36 Impact Attenuator/Crash Cushion, 37 Concrete Traffic Barrier, 38 Other Traffic Barrier, 39 Tree (Standing), 40 Hydrant, 41 Mailbox | - |
| **Collision with Bicycle or Moped** | 71 Riding in Bikeway<br>74 Riding off Roadway Direction | - |
| **Collision with MV in Transport (Except Moped)** | - | 80 Head-On,<br>83 Sideswipes Opposite, 85 Angle Opposite Direction |
| **Collision with MV - Other** | 102 Parked Motor Vehicle | - |

The independent variables included both continuous and indicator variables. For continuous variables, weighted averages were used. For some independent variables such as curvature, not only its mean but also its weighted standard deviation was calculated since it is reasonable to expect different crash frequencies in segments with identical curvature but with different standard deviations.

Other two important and unique features of the data synthesis process used in this study included the consideration of the direction of travel of the vehicle causing the crash and the general geometric environment of the roadway before the section or, equivalently, its alignment consistency. Separation of crashes by direction of travel allows the consideration of the different effects that variables such as grade may have on each direction for sections with steep grades. In addition, if the alignment consistency has an effect, the conditions prior to the section of interest can be quite different in each direction, something that cannot be accounted for when the crashes from both directions are grouped. Hence, the direction of all RwD crashes was extracted using the information in the motor vehicle crash reports and a new set of variables (for simplicity called environmental variables) were created, including the weighted average and the weighted standard deviation of curvature, grade, and IRI for 1.2-miles (~ 2.0 km) upstream of each segment. These were used to account for alignment consistency effects.

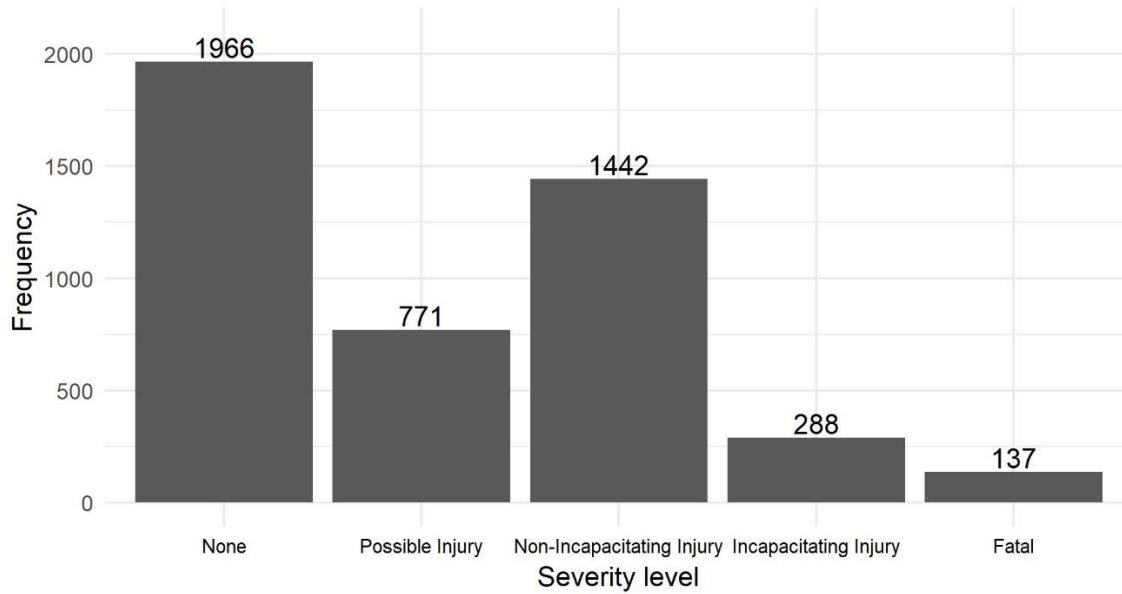

**FIGURE 1  Severity distribution of RwD crashes (none = property damage only).**

## Explanatory Variables

Two different categories of explanatory variables were used. The first category consists of three dummy or indicator variables including: a pavement type indicator = 0 if it is asphalt concrete and 1 otherwise, a shoulder type indicator = 0 if it is asphalt concrete and 1 otherwise, and a bridge indicator = 1 if there is a bridge in part of the section and 0 otherwise. The other category consists of the continuous explanatory variables shown in Table 2 along with some of their basic descriptive statistics. Most of the explanatory variables are self-explanatory (e.g., lane width, shoulder width, etc.) However, some of these deserve further explanation.

Since the models were developed with crashes for 10 years as a single observation for each section, the average AADT over those ten years was used as the AADT for the section. For most roadways in the state, the changes from one year to the next are very gradual and small. Thus, this simplification is minor when compared with the effect of the overall level of AADT of say, a road with about 1000 veh/day versus another with 20000 veh/day. AADT was used both as a measure of exposure by specifying log(AADT) as an offset and log(AADT) was also used as an explanatory variable to capture the effect that log(AADT) may have on that rate. A priori, it is expected that the rate of RwD crashes would tend to decrease with log(AADT), as the higher traffic interactions tend to make drivers more attentive with all else equal.

Since friction supply data were not available, the average friction demand on each segment was considered. This was estimated by solving for friction from the design curve equation using consistent units, resulting in:

$$f_s = \frac{V^2}{g\,R} - e \qquad (12)$$

where

$f_s$ = side friction demand,
R = radius of curvature,
$e$ = cross slope or superelevation in decimal (ft/ft or m/m),

$V$ = average traffic speed, and
$g$ = acceleration of gravity.

The average speed of the van used for collecting pavement distresses was used as a surrogate for the speed of traffic $V$ in the section. Consideration of speed through the side friction demand was useful for overcoming the problem of counterintuitive results often reported in the literature such as the speed limit being negatively associated with the crash frequency.

**TABLE 2  Descriptive statistics of the continuous variables**

| Variable | Min | Max | Mean | S.D. |
|---|---|---|---|---|
| AADT | 545 | 26682 | 6785 | 5895 |
| The Percentage of Single and Combination Trucks in the Stream | 0.00 | 56.57 | 5.59 | 7.30 |
| Average Side Friction Demand | 0.00 | 0.26 | 0.03 | 0.03 |
| Absolute Value of Curvature (degrees) | 0.00 | 52.28 | 3.43 | 6.69 |
| Standard Deviation of Curvature | 0.00 | 55.24 | 3.59 | 7.25 |
| Absolute Value of Curvature (Environmental Variable) | 0.00 | 35.57 | 3.36 | 5.95 |
| Standard Deviation of Curvature (Environmental Variable) | 0.00 | 41.46 | 4.57 | 7.15 |
| Grade (percent) | -12.89 | 12.07 | 0.00 | 3.11 |
| Standard Deviation of Grade | 0.00 | 6.68 | 0.49 | 0.66 |
| Grade (Environmental Variable) | 0.00 | 10.86 | 2.40 | 1.78 |
| Standard Deviation of Grade (Environmental Variable) | 0.00 | 7.38 | 1.30 | 0.94 |
| IRI (inch/mile) | 30.10 | 693.85 | 145.58 | 71.34 |
| Standard Deviation of IRI | 3.29 | 408.89 | 47.28 | 36.36 |
| IRI (Environmental Variable) | 34.40 | 586.74 | 144.75 | 65.19 |
| Standard Deviation of IRI (Environmental Variable) | 6.41 | 408.89 | 55.56 | 34.98 |
| Lane Width (feet) | 7.00 | 16.00 | 10.99 | 1.25 |
| Painted Median Width (feet) | 0.00 | 12.00 | 0.16 | 1.22 |
| Rutting (inch) | 0.00 | 0.67 | 0.06 | 0.05 |
| Shoulder Width (feet) | 0.00 | 18.00 | 4.81 | 3.04 |
| The Proportion of Total Length of sections with Guardrails (mile/mile) | 0.00 | 1.00 | 0.20 | 0.33 |
| The Proportion of Total Length of Sections with painted Medians (mile/mile) | 0.00 | 1.00 | 0.08 | 0.23 |
| The Proportion of Total Length of Sections with Asphalt Concrete Shoulder (mile/mile) | 0.00 | 1.00 | 0.90 | 0.40 |

The counterintuitive sign of the speed parameter estimate is typically because speed tends to be affected by the same unobserved factors affecting the crash rates, thus, causing the dependent variable and the error term to be correlated. This is a severe violation of the regression assumptions known as endogeneity that may lead to biased parameter estimates (even changing its sign, as it is apparently the case with speed). As shown later, the sign of the parameter estimate of side friction demand is intuitive and thus it helps to avoid dealing with the difficult problem of endogeneity.

Although all the selected roadways are TLTW highways, there are short segments with painted medians (usually to accommodate left-turning lanes/bays), which explains the use of variables related medians in TWTL highways.

Since the focus of this paper is not on the effects of individual variables but on the evaluation of the overall goodness of fit and because of space limitations, a more complete explanation of all explanatory variables will be provided elsewhere.

## ESTIMATION RESULTS

The three GLM methodologies described earlier were employed to develop crash frequency rate models with offsets for logs of $AADT_i$, $Length_i$, and $N_i$. Table 3 shows the parameter estimation results with R obtained using the package Mass (*23*) for the NB model, the package pscl (*24*) for the ZINB model, and the package lme4 (*25*) for the GLMM-NB model. In all cases, a 0.2-mile fixed section length was used.

Here, it suffices to mention that the estimation results were intuitively appealing in terms of the identification of statistically significant parameter estimates and their signs. As an example, it is seen that in the three models the parameter estimate of side friction demand is positive and statistically significant, indicating that as this variable increases within a section so do the probabilities of observing higher numbers of crashes. Space limitations preclude further discussion of the reasonability of many other interesting results, including the effects of the environmental or alignment consistency variables. These will be the topic of a separate paper.

Table 3 shows that the ZINB model results in a much larger number of statistically significant parameters than the NB model because of the zero-inflation portion of the model. On other hand, the GLMM-NB model results in fewer parameters. This is common with mixed-effects models since there is a tradeoff between the identification of fixed effects and variance components.

It is important to point out that the GLMM-NB model is presented here solely for illustrating the use of the proposed goodness of fit approach with different statistical methodologies as all the model estimations with this approach failed to converge because of nearly unidentifiability. Thus, a single random effect with each route used as the grouping variable was specified for the intercept term. The convergence failure is not surprising when using the section for grouping, as there is essentially a single observation per section. This makes nearly impossible the identification of the random effects since there is no information for identifying the unobserved heterogeneity of seemingly identical sections. This is the main reason why routes were used for grouping. The problem with this grouping, however, is that routes typically have very heterogeneous sections.

## PROPOSED GOODNESS OF FIT EVALUATION OF GLM MODELS

The literature on crash frequency models usually presents the "best model" selected based on some criterion such as the log-likelihood, the AIC, and the BIC. However, while these statistics are important for model selection, their values do not provide an easy interpretable comparison such as an $R^2$ in linear regression.

Unfortunately, unlike a regression model for a continuous dependent variable, a plot of deviance residuals is not very illuminating either.

**TABLE 3  Estimation results for the NB, ZINB, and GLMM-NB models**

| Parameter | NB model Estimate | NB model p-value | ZINB model Estimate | ZINB model p-value | GLMM-NB model Estimate | GLMM-NB model p-value |
|---|---|---|---|---|---|---|
| **Negative Binomial parameters** | | | | | | |
| Intercept | -5.88295 | < 2e-06 | -5.00938 | < 2e-06 | -7.20190 | < 2e-06 |
| log(AADT) | -0.37196 | < 2e-06 | -0.53604 | < 2e-06 | -0.24291 | 4.17e-09 |
| Mean Side Friction Demand | 3.00290 | 1.57e-07 | 2.46947 | 1.39e-05 | 3.53462 | 4.30e-09 |
| Grade (%) | -0.04166 | 1.93e-10 | -0.04112 | 1.36e-09 | -0.03430 | 1.22e-07 |
| (Grade)$^2$ | 0.00366 | 0.01395 | 0.00416 | 0.01105 | | |
| Absolute Value of Curvature | 0.08963 | < 2e-06 | 0.01011 | < 2e-06 | 0.10772 | < 2e-06 |
| (Curvature)$^2$ | -0.00214 | 8.62e-11 | 0.00034 | < 2e-06 | -0.00245 | 5.50e-15 |
| Lane Width (feet) | -0.04863 | 0.04669 | | | -0.05723 | 0.03266 |
| Shoulder Width (ft) | -0.04472 | 1.42e-07 | -0.05780 | 4.50e-11 | -0.05535 | 8.46e-11 |
| Average Environmental Curvature | -0.11247 | 9.84e-06 | -0.09213 | 0.00144 | -0.10700 | 0.00012 |
| Env. Curvature Standard Deviation | 0.06659 | 0.00170 | 0.09264 | 5.46e-05 | 0.09514 | 4.09e-05 |
| Percent Trucks | 0.01847 | 2.42e-06 | 0.01075 | 0.00765 | | |
| Prop. of Length with Median | -0.49253 | 2.07e-06 | -0.41544 | 3.61e-05 | -0.38550 | 5.86e-05 |
| Prop. of Length with Guardrails | -0.17273 | 0.00653 | -0.14720 | 0.01862 | | |
| Painted Median Width (feet) | -0.04783 | 0.02607 | -0.04494 | 0.03165 | | |
| Prop. Length with AC Shoulder | -0.18909 | 0.00189 | | | | |
| Average Environmental IRI | 0.00090 | 0.03170 | | | | |
| **Zero-inflation parameters** | | | | | | |
| Intercept | | | 8.20595 | 0.00022 | | |
| log(AADT) | | | -3.57741 | < 2e-06 | | |
| Absolute Value of Curvature | | | -0.05103 | 0.00856 | | |
| Percent Trucks | | | -0.12127 | 0.02800 | | |
| Lane Width (feet) | | | 0.95924 | 1.62e-08 | | |
| Shoulder Width (ft) | | | -0.29198 | 0.00035 | | |
| Average Environmental Curvature | | | 0.11063 | 4.50e-05 | | |
| Environmental Grade Std. Dev. | | | 0.39467 | 0.00026 | | |
| **Other parameters/statistics** | | | | | | |
| Theta ($\theta = 1/\alpha$) (Overdispersion) | 1.1068 | | 1.4038 | | 1.4322 | |
| Std. Dev. Of Route Random Effect | | | | | 0.5575 | |
| Akaike Information Criterion | 13707.0 | | 13568.3 | | 13386.1 | |
| Bayesian Information Criterion | 13827.2 | | 13735.3 | | 13473.1 | |
| Log-Likelihood | -6835.3 | | -6759.0 | | -6680.1 | |

Figure 2 shows the deviance residuals for the estimated NB model presented earlier. The bulk of the data corresponds to zero, one, or two crashes (darker areas), and the density of points decreases for more crashes since there are fewer observations.

Sometimes, the results are illustrated with a plot of estimated vs. observed values (*10*). However, what is typically presented as an observed value is the actual number of crashes observed on a given section, which is a particular realization of the random number of crashes that could have occurred in that section. On the other hand, what is presented as predicted is typically the predicted average number of crashes. Clearly, these two variables are not the same. This is also evidenced by what appears as a systematic bias in the predictions, as shown in Figure 3 for the NB model presented earlier. The problem with this approach is that in addition to comparing two different variables, it fails to recognize the probabilistic nature of the model being estimated.

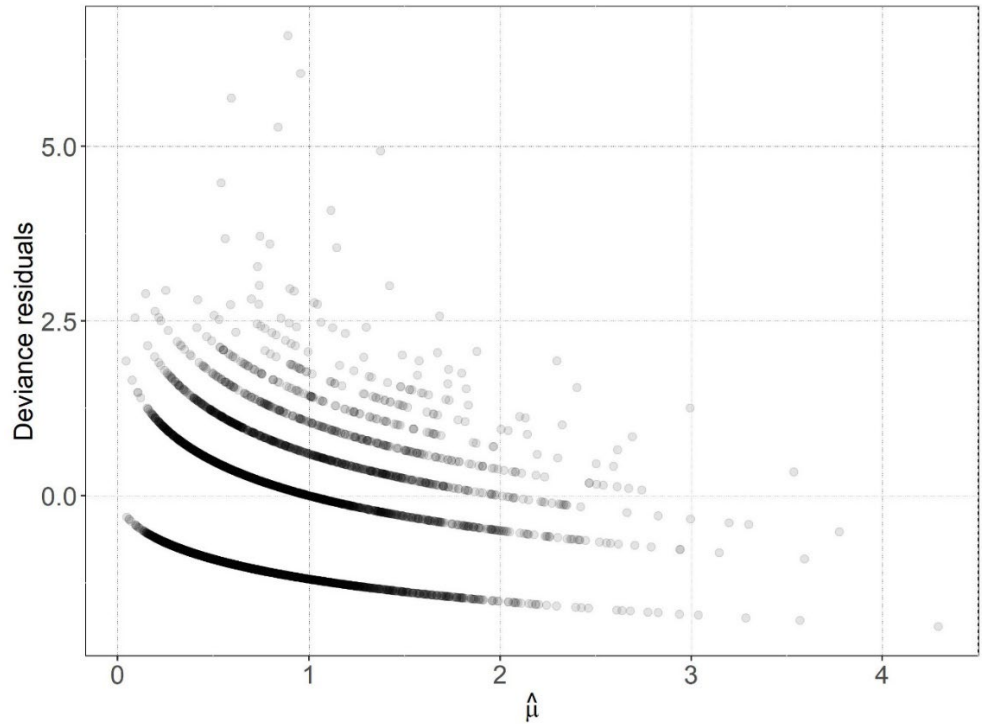

**FIGURE 2  Deviance residuals versus the average linear predictor ($\hat{\mu}$).**

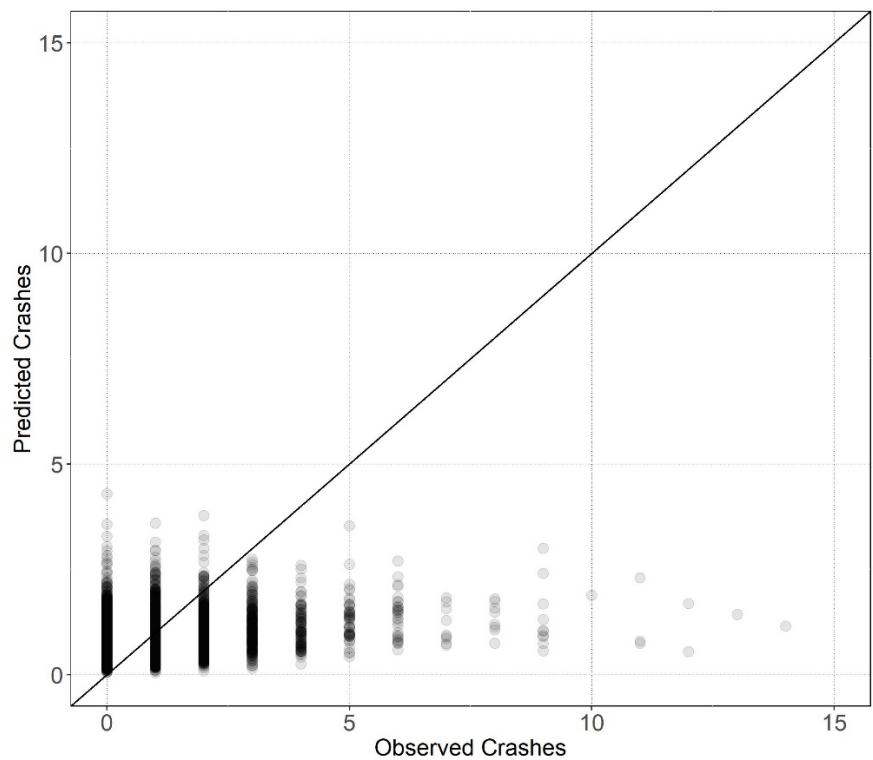

**FIGURE 3  Typical approach used to present observed vs. predicted crashes with observed crashes representing the particular realization of crashes on the section and the with predicted crashes representing the estimated average.**

Although crash frequency models are necessarily developed from observational studies where the researcher has no control over the independent variables, it is useful to think about a hypothetical experiment with many identical sections (such as the same curvature, traffic, grades, etc.) In such an experiment, no crashes would be observed for some sections; another set of sections would have only one crash, another would have two crashes and so on. If a model provides a good representation of the data generation process, then the observed and predicted proportions (or probabilities) should be similar. The problem is that the data in this study represent a cross-section of data, with a single observation for each section. Thus, in order to have replicates, one would need to identify nearly identical sections with a clustering algorithm. With so many explanatory variables, this becomes a non-trivial task.

Instead, the estimated models were used to predict the average number of crashes for each section and then the data were segmented based on these predictions. Sections were assigned to the different data segments defined by narrow intervals of the predicted average crash frequency (e.g., 0.0-0.2, 0.2-0.4, 0.4-0.6, etc.). Now, for a given data segment, the mid-value of the range was used to predict the probabilities of 0, 1, 2, … crashes in the segment. For example, for the data segment defined by the range 0.4-0.6, the value of $\lambda = 0.5$ was used to predict the probabilities. Those probabilities were then compared with the proportion of the sections assigned to the data segment on which 0, 1, 2, … crashes occurred.

The charts in Figure 4 compare the observed distributions (i.e., the observed proportions of sections with 0, 1, 2, … RwD crashes) with the distributions generated with the NB probability predictions for each data segment up to a maximum predicted average number of crashes of 2.0.

The black bars on each chart represent the empirical probability distribution of the number of RwD crashes generated with the sections for which the predicted average number of crashes from the estimated NB model fall within the narrow range for each chart in the figure. These are simply the proportions of the sections selected for each chart that have 0, 1, 2, … number of crashes. On the other hand, the gray bars illustrate the NB probability distribution predictions using the mid-value of the range of the average number of crashes corresponding to each chart.

The shape of the NB distribution can vary considerably for different ranges of its parameters (i.e., the average and the dispersion parameter). Therefore, it is essential to define relatively narrow ranges for the data segmentation.

The total number of sections whose expected number of RwD crashes falls within the corresponding chart range is provided on each chart together with the total number of crashes observed over the study period..

As shown in Figure 3, there are some sections with a predicted average number of crashes larger than 2 (more precisely, 156 sections for this NB model) but these are scattered over many data segments, resulting in very few observations for each segment and making it difficult to visualize the distribution. This is why no distributions are shown for any predicted average number of crashes above 2.

In general, the observed and predicted distributions are remarkably similar, especially for the ranges with more than 100 sections.

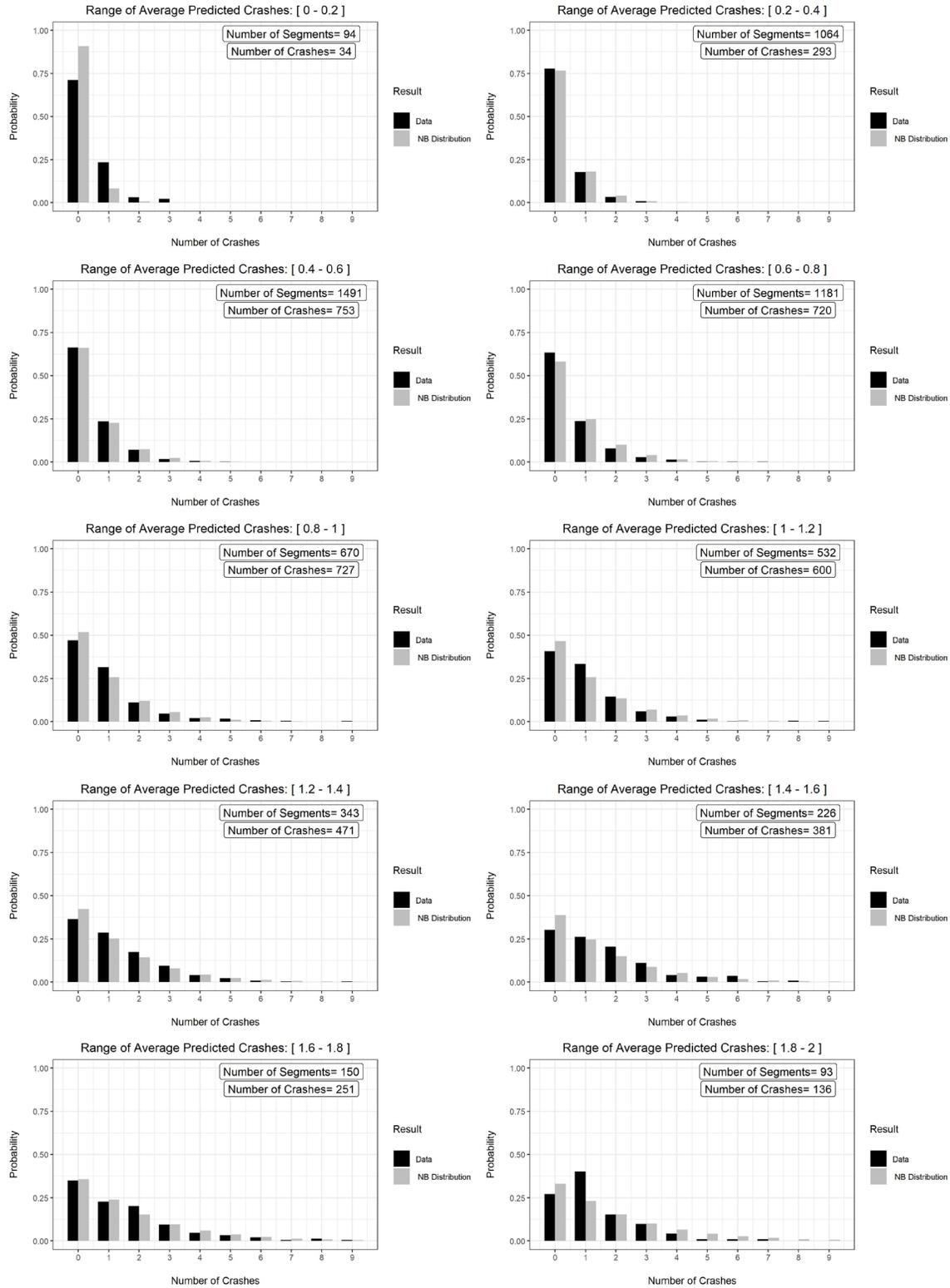

**FIGURE 4  Comparison graphs for small ranges of average predicted crashes.**

Figure 5a shows a weighted average of the distributions presented for each data segment in Figure 4, with the number of sections on each data segment used as weights. As can be seen, the observed and predicted distributions are remarkably similar. The charts in Figures 5b and 5c show the results obtained using the same procedure with the ZINB and GLMM-NB models. Because of the larger number parameters used for estimation, the ZINB model provides a slightly better fit than the NB model. This is barely noticeable in Figure 5, but it is a bit more noticeable with the charts for the individual data segments (not shown). On the other hand, the GLMM-NB model appears to over predict the proportion of sections with zero crashes and underpredict the proportions of 1 and 2 crashes (using only the fixed effects), which is not surprising given the fewer number of parameters identified as statistically significant. In addition, these results may have also been affected by the lack of convergence of the model because of the single observation per section issue discussed earlier, but experience of one of the authors with mixed-effects models for repeated measurement data of continuous variables indicates that in general, the more random parameters one includes in the model the lower is the fit with only fixed effects.

Figure 6 presents a way to summarize the information on each data segment with a single chart. In this figure, the average predicted and observed crash rates for each data segment are computed using the definition of the expected value of a discrete probability distribution, that is $E[X] = \sum_i f_i X_i$, where $f_i$ is the probability of observing $X_i$ crashes. In other words, each chart shown in Figure 4 becomes a point in this figure. With the generated points, it is easy to compute a pseudo $R^2$ value as:

$$R^2 = 1 - \frac{\sum_i N_i (O_i - P_i)^2}{\sum_i N_i (O_i - \bar{O})^2} \qquad (16)$$

where

$O_i$ = observed average number of sections in data segment $i$,
$\bar{O}$ = the average value of $O_i$,
$P_i$ = predicted average number of crashes in data segment $i$, and
$N_i$ = number of observations in data segment $i$.

This is basically the normal $R^2$ of linear regression with minor modifications. One is that the sums of squares are weighted by the number of sections defining each data point to avoid giving too much weight to data segments with only a few sections. Note that for a perfect model represented by the 45º line, a value of $P_i$ in the predicted axis would result in the same value in the observed axis. Thus, $P_i$ is taken as the model prediction for the same value in the other axis and thus, the value of $(O_i - P_i)$ measures the error from that prediction. The denominator in Eq. 16 is basically a weighted total sum of squares.

In addition to the maximum log-likelihood value, the AIC and BIC statistics were utilized to evaluate the results of fitted models. AIC and BIC deal with the trade-off between the goodness of fit of the model and the simplicity of the model by introducing a penalty term for the number of parameters in the model. In general, a model with lower AIC and BIC values or a larger LL values is preferred. As shown in Table 3, in this regard, the GLMM-NB model is better than the ZINB model and, in turn, the ZINB model is better than the NB model. The better statistics for the GLMM-NB model are the result of its flexibility to predict, in theory, a separate intercept for each section. However, it must be remembered that the GLMM-NB model did not converge properly. The statistical superiority of the ZINB model over the NB model was also confirmed by a Vuong test.

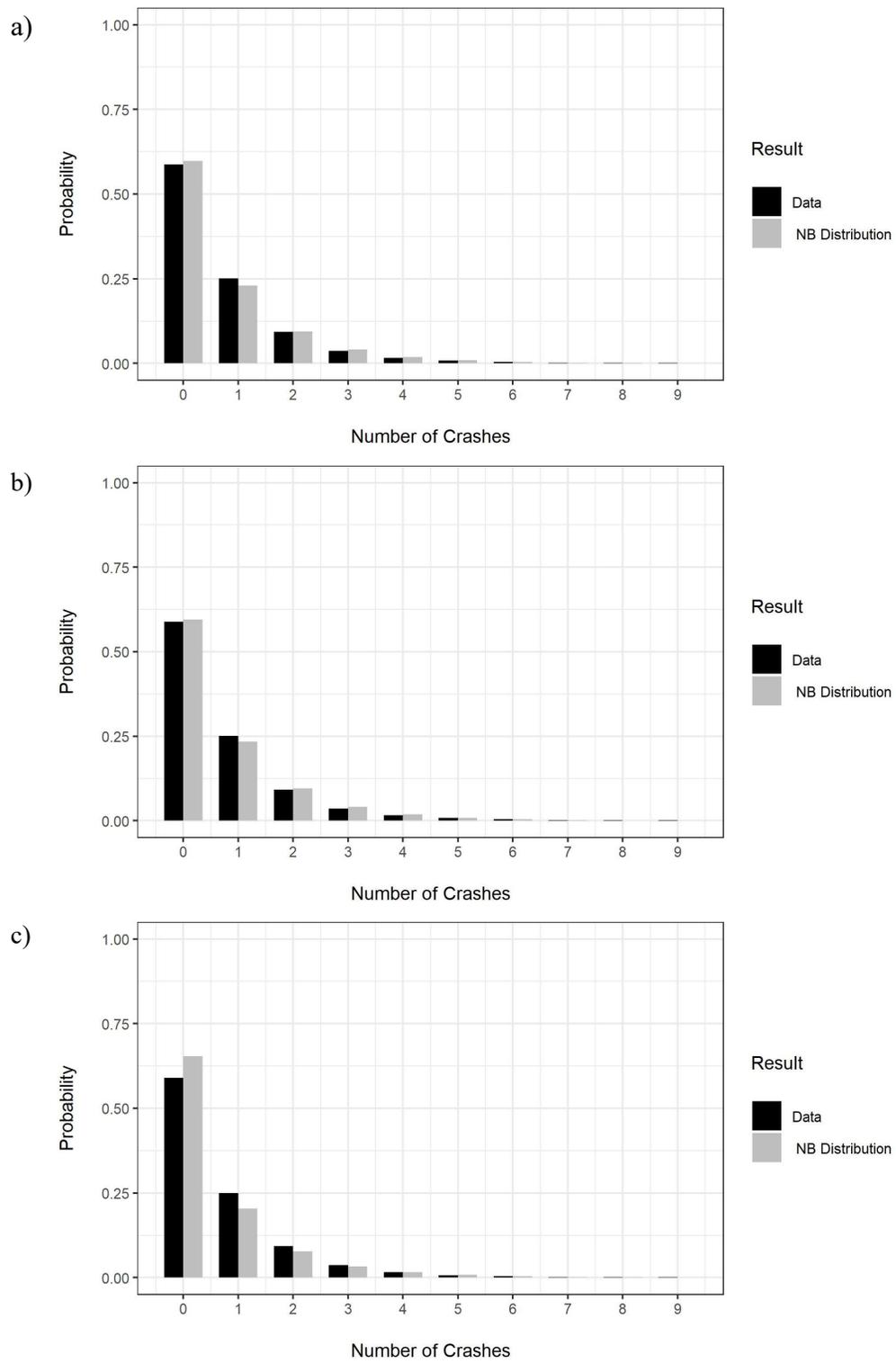

**FIGURE 5  Weighted average of the distributions for all the data segments a) NB, b) ZINB, and c) GLMM-NB**.

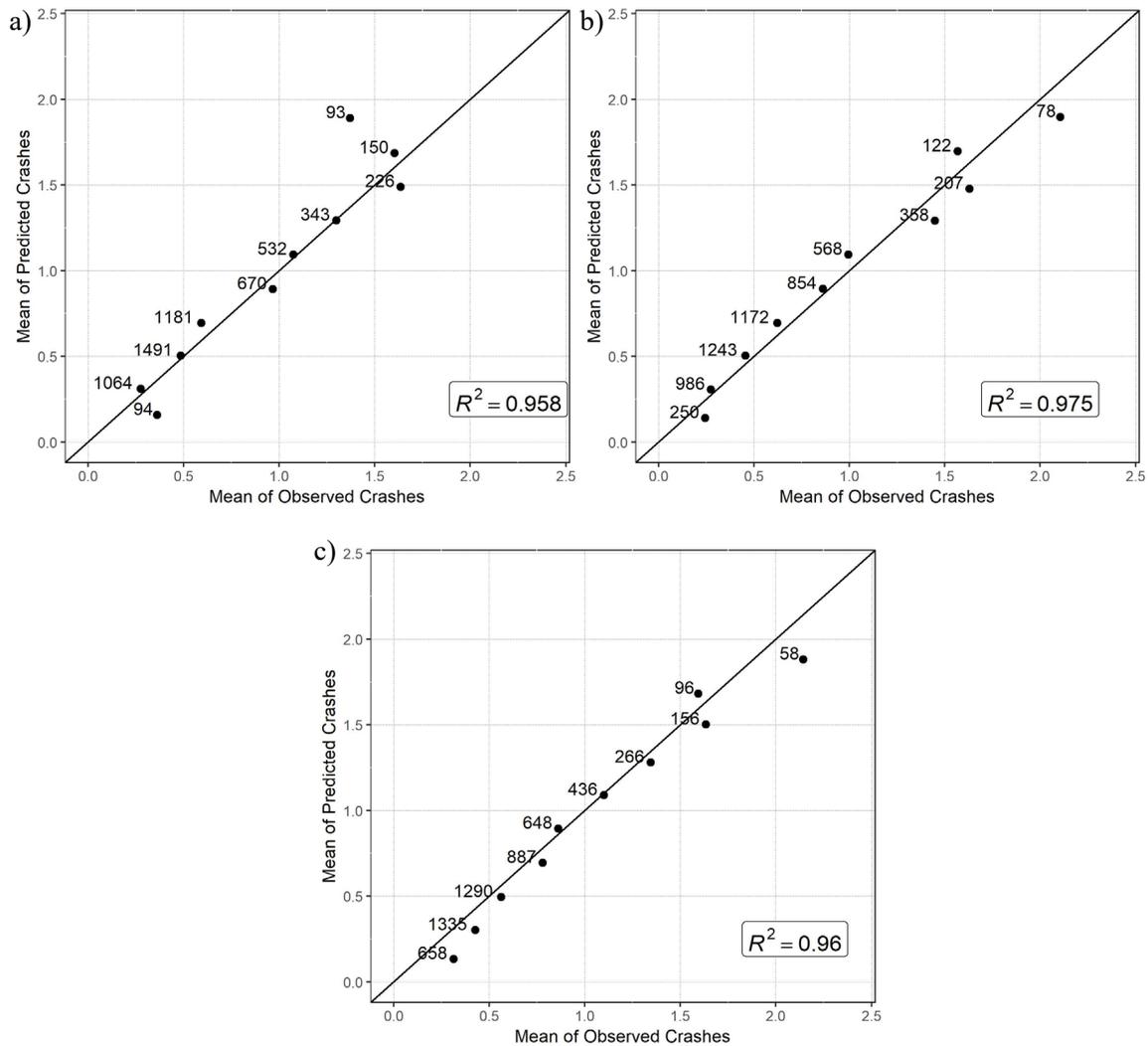

**FIGURE 6  Mean of observed versus mean of predicted crashes a) NB, b) ZINB, and c) GLMM-NB**.

As shown in Figure 6a for the NB model, the pseudo $R^2$ is 0.958. This provides confirmation of the quality of the fitted model. It should be noted that the same computations without weights provided too much importance to the points defined with very few sections, resulting in a value of about 0.85.

Figure 6b confirms the closer fit obtained with the ZINB model. The higher $R^2 = 0.975$ is consistent with the better statistics obtained for the ZINB when compared with the NB model (see Table 3).

Figure 6c shows the fit with the GLMM-NB model. In this case, the points are less evenly distributed on each side of the line. A careful inspection shows a slight curvilinear trend. Nevertheless, the fit is still very good with a pseudo $R^2 = 0.960$. Once again, it is important to remember that the GLMM-NB model did not converge. Note that the lower $R^2$, when compared with the ZINB model, is not surprising despite the better AIC, BIC, and LL values of the GLMM-NB model. For a better representation of the data generation process, the GLMM-NB model makes some trade-offs between identification of fixed-effects and the identification of variance parameters of the random effects. While this provides a better statistical model (better representation of the data generation process), the variance parameters of the random effects are not used for prediction. Hence, a lower pseudo $R^2$ should generally be expected for the GLMM-NB models.

One difficult aspect of the estimation of mixed-effects models is the determination of how many random effects are appropriate. The data limitations in this study did not allow further exploration of this issue but this is an area where the use of the plots and pseudo $R^2$ proposed in the study could be extremely helpful. Specifically, they can help the researcher determine the point at which addition of more random-effects would result in lower quality fits despite improvements in other statistics.

Each approach used for model estimation is based on different assumptions about the data generation process, but the procedure presented in this section indicates that all three models provided good fits.

Regarding the simplicity in the interpretation of results, the ZINB model is more challenging than the NB regression. The mixed-effects negative binomial model (the most complex methodology for estimation) had the best statistics of the three models (log-likelihood, AIC, and BIC).

Consideration of the assumptions behind each methodology is also crucial. While assuming that a set of observations that are necessarily zeroes in the ZINB model is defensible in other fields of science, it is less so in crash frequency modeling since no section is inherently safe.

A common challenge with the mixed-effects methodology is the uncertainty that exists in the interpretation of results. In numerous studies (*10, 16*), the estimated standard deviation of a random effect is close to or even higher than the corresponding estimated fixed-effects coefficient. Therefore, the interpretation of results for such parameters remains inconclusive since they would suggest that a unit increase in a variable would increase the number of crashes for a certain proportion of the sections and it would reduce the number of crashes for the complementary proportion of sections. The authors suspect that despite the better statistics, some of these results are obtained at the expense of considerably lower fits with only the fixed effects. The proposed procedure may be helpful for identifying a parsimonious model with only some random effects that balance the desire to model the data generation process adequately, so the statistics and hypothesis test are valid, versus the goodness-of-fit required in practice for prediction.

## CONCLUSIONS

This paper presented a new procedure for evaluating the goodness of fit of GLM models estimated with RwD crash frequency for the State of Hawaii on TLTW state roads. The analyses were based on ten years of RwD crashes data (including all severity levels) and roadway characteristics (e.g., traffic, geometry, and inventory databases) that can be aggregated at the section level. The proposed procedure complements other statistics such as the AIC, BIC, and LL used for model selection. It is consistent with those statistics for models without random effects, but it diverges for GLMM-NB models. The procedure can be useful for selecting an appropriate number of random effects in mixed-effects models without losing much of the fit of the model.

The procedure also provides a clear visualization of the fit of crash frequency models and allows the computation of a pseudo $R^2$ similar to the one used in linear regression, which is a complement to existing procedures. The NB, ZINB, and GLMM-NB models estimated for predicting the frequency of RwD crashes in TLTW state roads in Hawaii provide reasonably robust approximations to the distribution of RwD crashes in the state.

It is recommended to evaluate the use of the procedure for selecting an appropriate number of random-effects in GLMM-NB models that balance the consideration of unobserved heterogeneity and the goodness-of-fit by using more appropriate datasets that do not lead to convergence problems. Exploration of different weighting schemes for the proposed pseudo $R^2$ is also desirable.

## ACKNOWLEDGMENTS

The financial support of the State of Hawaii Department of Transportation (HDOT) is greatly appreciated and acknowledged. The contents of this paper reflect the view of the writers, who are responsible for the facts and accuracy of the data presented herein. The contents do not necessarily reflect the official views or policies of the State of Hawaii Department of Transportation or the Federal Highway Administration. The contents contained herein do not constitute a standard, specification, or regulation.

## REFERENCES

1) Federal Highway Administration. Roadway Departure (RwD) Strategic Plan. https://safety.fhwa.dot.gov/roadway_dept/docs/rwd_strategic_plan_version2013.pdf. Accessed Mar. 1, 2018.
2) National Highway Traffic Safety Administration. Fatality Analysis Reporting System (FARS). https://www.nhtsa.gov/research-data/fatality-analysis-reporting-system-fars. Accessed July 29, 2019.
3) Mannering, F. L., and C. R. Bhat. Analytic methods in accident research: Methodological frontier and future directions. *Analytic Methods in Accident Research*, Vol. 1, Jan. 2014, pp. 1–22.
4) Miaou, S.-P. The relationship between truck accidents and geometric design of road sections: Poisson versus negative binomial regressions. *Accident Analysis & Prevention*, Vol. 26, No. 4, 1994, pp. 471–482.
5) Shankar, V., F. Mannering, and W. Barfield. Effect of roadway geometrics and environmental factors on rural freeway accident frequencies. *Accident Analysis & Prevention*, Vol. 27, No. 3, Jun. 1995, pp. 371–389.
6) Lee, J., and F. Mannering. Impact of roadside features on the frequency and severity of run-off-roadway accidents: an empirical analysis. *Accident Analysis & Prevention*, Vol. 34, No. 2, Mar. 2002, pp. 149–161.
7) Lord, D., S. D. Guikema, and S. R. Geedipally. Application of the Conway-Maxwell-Poisson generalized linear model for analyzing motor vehicle crashes. *Accident Analysis and Prevention*, Vol. 40, No. 3, May 2008, pp. 1123–1134.
8) Hauer, E. *The art of regression modeling in road safety*. Springer, 2016.
9) Geedipally, S. R., D. Lord, and S. S. Dhavala. The negative binomial-Lindley generalized linear model: Characteristics and application using crash data. *Accident Analysis and Prevention*, Vol. 45, Mar. 2012, pp. 258–265.

10) Anastasopoulos, P. C., and F. L. Mannering. A note on modeling vehicle accident frequencies with random-parameters count models. *Accident Analysis and Prevention*, Vol. 41, No. 1, Jan. 2009, pp. 153–9.

11) Faraway, J. J. *Extending the linear model with R: generalized linear, mixed effects and nonparametric regression models*. CRC press, 2016.

12) Mannering, F. L., V. Shankar, and C. R. Bhat. Unobserved heterogeneity and the statistical analysis of highway accident data. *Analytic Methods in Accident Research*, Vol. 11, Sep. 2016, pp. 1–16.

13) Agresti, A. *Foundations of linear and generalized linear models*. John Wiley & Sons, 2015.

14) Lord, D., and F. Mannering. The statistical analysis of crash-frequency data: A review and assessment of methodological alternatives. *Transportation Research Part A: Policy and Practice*, Vol. 44, No. 5, Jun. 2010, pp. 291–305.

15) Washington, S. P., M. G. Karlaftis, and F. L. Mannering. *Statistical and econometric methods for transportation data analysis*. CRC press, 2010.

16) Winkelmann, R. *Econometric analysis of count data.* Springer Science & Business Media, 2008.

17) Lord, D., S. P. Washington, and J. N. Ivan. Poisson, poisson-gamma and zero-inflated regression models of motor vehicle crashes: Balancing statistical fit and theory. *Accident Analysis and Prevention*, Vol. 37, No. 1, Jan. 2005, pp. 35–46.

18) Lord, D., S. Washington, and J. N. Ivan. Further notes on the application of zero-inflated models in highway safety. *Accident Analysis and Prevention*, Vol. 39, No. 1, Jan. 2007, pp. 53–57.

19) Venkataraman, N., G. F. Ulfarsson, and V. N. Shankar. Random parameter models of interstate crash frequencies by severity, number of vehicles involved, collision and location type. *Accident Analysis & Prevention*, Vol. 59, 2013, pp. 309–318.

20) Zuur, A. F., E. N. Ieno, N. J. Walker, A. A. Saveliev, and G. M. Smith. *Mixed effects models and extensions in ecology with R.* Gail M, Krickeberg K, Samet JM, Tsiatis A, Wong W, editors. New York, NY: Spring Science and Business Media, 2009.

21) Pinheiro, J. C., and D. M. Bates. Mixed-Effects Models in S and S-plus. Statistics and computing, 1978.

22) R Core Team. R: A Language and Environment for Statistical Computing. http://www.r-project.org/. Accessed July 29, 2019.

23) Ripley, B. *MASS: Support Functions and Datasets for Venables and Ripley's MASS.* https://cran.r-project.org/package=MASS. Accessed July 29, 2019.

24) Fearon, S. J. and A. T. and A. Z. and C. M. and J. *pscl: Political Science Computational Laboratory.* https://cran.r-project.org/package=pscl. Accessed July 29, 2019.

25) Walker, D. B. and M. M. and B. B. and S. *lme4: Linear Mixed-Effects Models using "Eigen" and S4.* https://cran.r-project.org/package=lme4. Accessed July 29, 2019.

26) Rusli, R., M. M. Haque, M. King, and W. S. Voon. Single-vehicle crashes along rural mountainous highways in Malaysia: An application of random parameters negative binomial model. *Accident Analysis & Prevention*, Vol. 102, 2017, pp. 153–164.